\documentstyle{article}

\def \lket {|}
\def \rket {\rangle}
\def \lbra {\langle}

\newcommand{\ket}[1]{\lket #1\rket}

\newtheorem{Lemma}{Lemma}

\newcommand{\proof}{\noindent {\bf Proof: }}
\newcommand{\qed}{$\Box$}
\def\H{{\cal H}}

\begin{document}
\title{Lower bound for a class of weak quantum coin flipping protocols}

\date{}

\author{Andris Ambainis\thanks{Supported
  by NSF Grant CCR-9987845 and the State of
  New Jersey.}\\
       School of Mathematics\\
       Institute for Advanced Study\\
       Princeton, NJ 08540\\
       e-mail: {\tt ambainis@ias.edu}}
\maketitle 

\begin{abstract}
We study the class of protocols for weak quantum coin flipping
introduced by Spekkens and Rudolph (quant-ph/0202118).
We show that, for any protocol in this class,
one party can win the coin flip with probability at least 
$1/\sqrt{2}$.
\end{abstract}

\section{Introduction}

Coin flipping is a cryptographic primitive in which
two parties (Alice and Bob) together generate
a random bit so that the value of the random bit cannot
be controlled by any one party. If both parties are honest, the random 
bit must be 0 with probability 1/2 and 1 with probability 1/2.
If one party is honest but the other is not, the honest
party is still guaranteed that the cheater cannot
control the outcome.

There are two variants of this requirement. 
In {\em strong coin flipping}, we require that,
no matter what a dishonest Alice (dishonest Bob) does, 
the probability of the result being $a$ is at most $P_A$ 
(at most $P_B$), for each of the two possible outcomes
$a\in\{0, 1\}$. In {\em weak coin flipping},
we know in advance that one outcome (say, 0) benefits
Alice and the other outcome (say, 1) benefits Bob.
Therefore, we only require that dishonest Alice cannot make
result 0 with probability more than $P_A$ and
dishonest Bob cannot make the result 1 with 
probability more than $P_B$.

Coin flipping is possible classically with complexity
assumptions such as the existence of one-way functions \cite{Blum}.
In an information-theoretic setting (parties with
unlimited computational power), in any classical protocol
there is a party which can set the outcome to 0 with certainty and
1 with certainty. Thus, neither of the two variants
is possible classically information-theoretically. 

In the quantum model, strong coin flipping has been studied 
by \cite{LC,MSC,ATVY,Ambainis,RS0,Leslau,Kitaev}. 
The best protocol \cite{Ambainis,RS0} can achieve
any combination of $P_A$ and $P_B$ such that $0\leq P_A$, $0\leq P_B$, 
$P_A+P_B=\frac{1}{2}$. In particular, if we want to have
the same security guarantees for both parties,
we can achieve $P_A=P_B=\frac{3}{4}$.
The best known lower bound is 
that any protocol for strong coin flipping
must have $P_A P_B\geq \frac{1}{2}$ \cite{Kitaev}.
If we want to have
the same security guarantees for both parties,
this gives $P_A=P_B=\frac{1}{\sqrt{2}}$.
This is quite close to what is achieved by \cite{Ambainis,RS0}.

Less is known about weak coin flipping.
The lower bound of \cite{Kitaev} does not apply
to weak coin flipping. Thus, we might still have
a protocol for weak coin flipping with 
$P_A=\frac{1}{2}+\epsilon$ and $P_B=\frac{1}{2}+\epsilon$
for an arbitrarily small $\epsilon>0$.
(An ``exact'' protocol with $\epsilon=0$ 
is impossible because the impossibility proof
for exact protocols from \cite{MSC} applies to weak coin flipping.
Also, we know that if $\epsilon>0$ is achievable, 
at least $\Omega(\log\log \frac{1}{\epsilon})$ rounds are needed \cite{Ambainis}.)

Weak coin flipping has been studied
by \cite{Goldenberg,RS}.
The first protocol \cite{Goldenberg}
achieved $P_A=P_B\approx 0.327...$.
\cite{RS} described a general class of protocols
and showed that this class achieves any combination of 
$P_A$, $P_B$ such that $0< P_A\leq 1$, $0< P_B\leq 1$ and 
$P_A P_B=\frac{1}{2}$.
(The protocol achieving $P_A P_B=\frac{1}{2}$ was
also independently discovered by the author of this note.)
\cite{RS} conjectured that this is the best possible
for this class of protocols.
In this note, we prove this conjecture.

\section{A class of protocols}

Rudolph and Spekkens \cite{RS} considered the following
class of protocols for weak coin flipping:

\begin{enumerate}
\item
Alice prepares a pair of systems in a state 
$\ket{\psi}\in \H_A \otimes \H_B$ and sends 
the system $B$ to Bob.
\item
Bob performs the POVM measurement $\{ E_0, E_1\}$
on $\H_B$, sends a classical bit $b$ with the outcome
of the measurement to Alice.
\item
If $b=0$, Bob sends the system $B$ back to Alice.
If $b=1$, Alice sends the system $A$ to Bob.
The party that receives the system then checks
whether the joint state of $A$ and $B$ is
$\ket{\psi_b}=\frac{I\otimes \sqrt{E_b}\ket{\psi}}{
\sqrt{\lbra \psi | I\otimes E_b | \psi\rket}}$
by measuring an observable consisting
of $\ket{\psi_b}$ and its orthogonal complement.
The possibilities are
\begin{enumerate}
\item
$b=0$, Alice finds $\ket{\psi_0}$. Bob wins.
\item
$b=0$, Alice does not find $\ket{\psi_0}$. Alice has caught
Bob cheating.
\item
$b=1$, Bob finds $\ket{\psi_1}$. Alice wins.
\item
$b=1$, Bob does not find $\ket{\psi_1}$. Bob has caught
Alice cheating.
\end{enumerate}
\end{enumerate}

Different choices of $\ket{\psi}$, $E_0$ and $E_1$ give
different protocols.
\cite{RS} showed how to achieve any combination
of $0<P_A\leq 1$, $0<P_B\leq 1$ such that $P_A P_B=\frac{1}{2}$.
They also showed that this is the best possible for
this protocol using two-dimensional systems $\H_A$ 
and $\H_B$ and conjectured that this is the best 
for $\H_A$ and $\H_B$ of arbitrary dimension.
Thus, using qubits would be optimal for this class
of protocols, unlike in the known protocols for
strong quantum coin flipping \cite{Ambainis,RS0}
where qutrits are needed to achieve the best results.

\section{The lower bound}

We now prove this conjecture.

Let $P_A^{max}(E_0, \ket{\psi})$ and
$P_B^{max}(E_0, \ket{\psi})$ be the
maximum probabilities of winning for Alice and Bob, 
for the given choices of $E_0$ and $\ket{\psi}$.
We use the expressions for $P_A^{max}$ and $P_B^{max}$ shown
by \cite{RS}: 
\[ P^{max}_A = 2 Tr(\rho E^2_0) , \mbox{~~~~} P^{max}_B = 2 Tr(\sqrt{\rho E_0 \rho})^2, \]
where $\rho$ is the density matrix of Bob's part of $\ket{\psi}$.
We will show that the product of these expressions is at least $\frac{1}{2}$.
The first step is to show that it is enough to consider the case 
when the Schmidt decomposition of $\ket{\psi}$ consists
of eigenvectors of $E_0$.

\begin{Lemma}
For any choice
of $\ket{\psi}$ and $E_0$ in the protocol of \cite{RS}, 
there exists $\ket{\tilde{\psi}}$ 
such that Bob's part of
Schmidt decomposition of $\ket{\tilde{\psi}}$ consists
of eigenvectors of $E_0$
and $P_A^{max}(E_0, \ket{\tilde{\psi}})\leq P_A^{max}(E_0, \ket{\psi})$,
$P_B^{max}(E_0, \ket{\tilde{\psi}})\leq P_B^{max}(E_0, \ket{\psi})$.
\end{Lemma}

\proof
Let $\ket{\phi_1}$, $\ldots$, $\ket{\phi_k}$ be the eigenvectors
of $E_0$. Since $E_1=I-E_0$, they are also eigenvectors of $E_1$.
We write the state $\ket{\psi}$ sent by Alice
in round 1 as
\[ \ket{\psi}=\sum_{i=1}^k \lambda_i \ket{\varphi_i}\ket{\phi_i} .\] 
Notice that this is not a Schmidt decomposition because
$\ket{\phi_i}$ are not necessarily orthogonal. We consider a protocol
in which Alice sends the state
\[ \ket{\tilde{\psi}}=\sum_{i=1}^k \lambda_i \ket{i}\ket{\phi_i} \]
instead of $\ket{\psi}$. We claim that 
$P_A^{max}$ and $P_B^{max}$ in this protocol are less than or equal to 
$P_A^{max}$ and $P_B^{max}$ when Alice sends $\ket{\psi}$.
This is shown by mapping Alice's and Bob's cheating strategies from
the protocol with $\ket{\tilde{\psi}}$ to the protocol with $\ket{\psi}$.

{\bf Case 1: Alice.}
The most general strategy of Alice is to prepare a state
\[ \ket{\psi'}=\sum_{i=1}^k \mu_i \ket{\varphi'_i} \ket{\phi_i} .\]
Bob's measurement splits the state into two parts $\ket{\psi'_0}$ 
and $\ket{\psi'_1}$. Since $\ket{\phi_i}$ are eigenvectors of $E_0$ 
and $E_1$, 
\[ \ket{\psi'_1}= \sum_{i=1}^k \mu'_i \ket{\varphi'_i} \ket{\phi_i} .\]
After Alice sending her part to Bob, Bob tests the state $\ket{\psi'_1}$
against the state $\ket{\tilde{\psi_1}}$ which
would have resulted if Alice had prepared the honest state 
$\ket{\tilde{\psi}}$.
Since $\ket{\tilde{\psi}}$ is a superposition of $\ket{i}\ket{\phi_i}$ and
$E_0$, $E_1$ are diagonal in the basis consisting of $\ket{\phi_i}$,
$\ket{\tilde{\psi_1}}$ is a superposition of $\ket{i}\ket{\phi_i}$
as well. Therefore, the inner product between $\ket{\psi'_1}$
and $\ket{\psi_1}$ is maximized if $\ket{\varphi'_i}=\ket{i}$
for all $i$ and, if Alice sends 
\[ \ket{\psi''}=\sum_{i=1}^k \mu_i \ket{i} \ket{\phi_i} \]
instead of $\ket{\psi'}$, this only increases her success probability.
To finish the proof, notice that sending the state 
\[ \ket{\psi'''}=\sum_{i=1}^k \mu_i \ket{\varphi_i} \ket{\phi_i} \]
in the protocol for $\ket{\psi}$ achieves the same probability as
sending $\ket{\psi''}$ in the protocol for $\ket{\tilde{\psi}}$. 

{\bf Case 2: Bob.}
An honest Bob's measurement splits $\ket{\tilde{\psi}}$ into states 
$\ket{\tilde{\psi_0}}$ and $\ket{\tilde{\psi_1}}$.
Since $E_0$ and $E_1$ are diagonal in the basis $\ket{\phi_i}$,
the state $\ket{\tilde{\psi_0}}$ is of the form
\[ \ket{\tilde{\psi_0}}= \sum_i a_i \ket{i}\ket{\phi_i}. \] 

A dishonest Bob's most general strategy is to transform the
state $\ket{\tilde{\psi}}$ into a state having maximum overlap
with $\ket{\tilde{\psi_0}}$. Since he cannot access $\ket{i}$,
the state having maximum overlap is just $\ket{\tilde{\psi}}$. Therefore,
Bob's best strategy is just to leave $\ket{\tilde{\psi}}$ unchanged, claim
$b=0$ and send his part of the state back to Alice.
The same success probability can be achieved by Bob in the protocol 
for $\ket{\psi}$ by a similar strategy (claim $b=0$ and send the
state back).
\qed

Similarly to \cite{RS}, let $\rho$ be the density matrix of
Bob's side of $\ket{\psi}$.
We write density matrices $\rho$ and $E_0$ in the basis
consisting of $\ket{\psi_i}$. 
Both matrices are diagonal in this basis.
Let $a_i$ be the elements on the diagonal of $\rho$
and $b_i$ be the elements on the diagonal of $E_0$.
Then,
\[ P^{max}_A = 2 Tr(\rho E^2_0) = 2 \sum_{i=1}^k a_i b_i^2 ,\]
\[ P^{max}_B = 2 Tr(\sqrt{\rho E_0 \rho})^2 
= 2 (\sum_{i=1}^2 a_i \sqrt{b_i})^2 \]
and we have the extra constraint that $Tr(\rho E_0)=\sum_{i} a_i b_i = 
\frac{1}{2}$
(because the outcome of an honest coin flip is 0 with probability 1/2).

By Holder's inequality, we have  
$\|x\|_3 \|y\|_{\frac{3}{2}}\geq \lbra x | y \rket$
and $\|x\|^3_3 \|y\|^3_{\frac{3}{2}}\geq \lbra x | y \rket^3 $
for any vectors $x$, $y$. 
Applying this inequality to 
$x=(a^{1/3}_i b^{2/3}_i)_{i=1}^k$ and  
$y=(a^{2/3}_i b^{1/3}_i)_{i=1}^k$ gives us
\[  P^{max}_A P^{max}_B = 4 \sum_{i=1}^k a_i b_i^2 
\left( \sum_{i=1}^k a_i \sqrt{b_i}\right)^2 \geq 
4 \left(\sum_{i=1}^k a_i b_i\right)^3 = 4 \left( \frac{1}{2} \right)^3 = 
\frac{1}{2} .\]

\section{Conclusion}

We have shown that any choice of parameters in
the protocol of \cite{RS} gives 
$P_A^{max} P_B^{max} \geq \frac{1}{2}$.
Curiously, this is the same as the lower bound
of \cite{Kitaev} for arbitrary protocols for
{\em strong coin flipping}.
However, there does not seem to be any direct 
connection between the two results.
It remains open whether a different protocol
(not in the class described above)
for weak coin flipping could achieve a better security.

Another interesting question about coin flipping
protocols is ``cheat-sensitivity'' studied by \cite{ATVY,Kent,RS}.
A protocol for coin flipping or other cryptographic tasks
is cheat-sensitive if a dishonest party cannot
increase the probability of one outcome without
being detected with some probability. 
Many quantum protocols display some cheat-sensitivity
but it remains to be seen what degree of cheat-sensitivity
can be achieved.

{\bf Acknowledgment.}
Thanks to Terry Rudolph for useful comments.

\end{document}